\documentclass[11pt]{article}

\usepackage{amsmath, amssymb}

\usepackage{graphicx}

\usepackage{geometry}
\usepackage[numbers]{natbib}

\usepackage[hyperfootnotes=false, colorlinks=true, linkcolor=blue, citecolor=blue, urlcolor=blue]{hyperref} 
\usepackage[nameinlink]{cleveref} 

\usepackage{xurl}

\newcommand{\figref}[1]{\cref{#1}}

\title{\Large\itshape The Potential of Nighttime Light Imagery \\ for Detailed Local Economic Analysis}
\author{Shoichi Otomo}
\date{}

\begin{document}

\maketitle

\begin{abstract}

This manuscript is an English translation and extended version of a paper originally published in Japanese~\cite{otomo2}.
Driven by remarkable advances in remote sensing and big data processing, spatial technologies are increasingly leveraged in economic research. 
While satellite nighttime light intensity is widely recognized for tracking macroeconomic parameters—such as regional GDP, employment, and population—less attention has been paid to fine-grained spatial processing methodologies for local tourism economies. 
This study first details a raster processing technique applied to nightlight imagery. 
It then focuses on Yuzawa Town (Uonuma District, Niigata Prefecture), analyzing the local economy and sports/tourist attractions. 
Specifically, I investigate the spatio-temporal relationships between tourist arrivals and various local statistical datasets. 
Furthermore, this paper highlights how nightlight data can capture intra-municipal economic dynamics that remain undetectable through standard macroeconomic indicators.

\end{abstract}


\section{Introduction}
\subsection{Background}
Driven by recent progress in geospatial computing and open data frameworks, high-resolution satellite remote sensing has increasingly been adopted for sub-national economic analysis. While macroeconomic statistics remain the standard tool for regional assessment, they often lack temporal frequency and finer spatial resolution. This limitation is particularly critical in developing regions or specialized local economies where official administrative statistics are updated infrequently or lack granularity. Conventional macroeconomic indicators such as GDP also face inherent measurement challenges, including currency fluctuations and unrecorded seasonal economic shifts.

To complement traditional data, satellite-derived nighttime light (NTL) observation offers a standardized, continuous metric for capturing spatio-temporal dynamics in local human activity. Because NTL data allow for consistent regional comparisons without relying on localized reporting standards, their application has expanded into regional policy planning and tourism management.

\begin{figure}[h]
\begin{center}
\includegraphics[width=130mm]{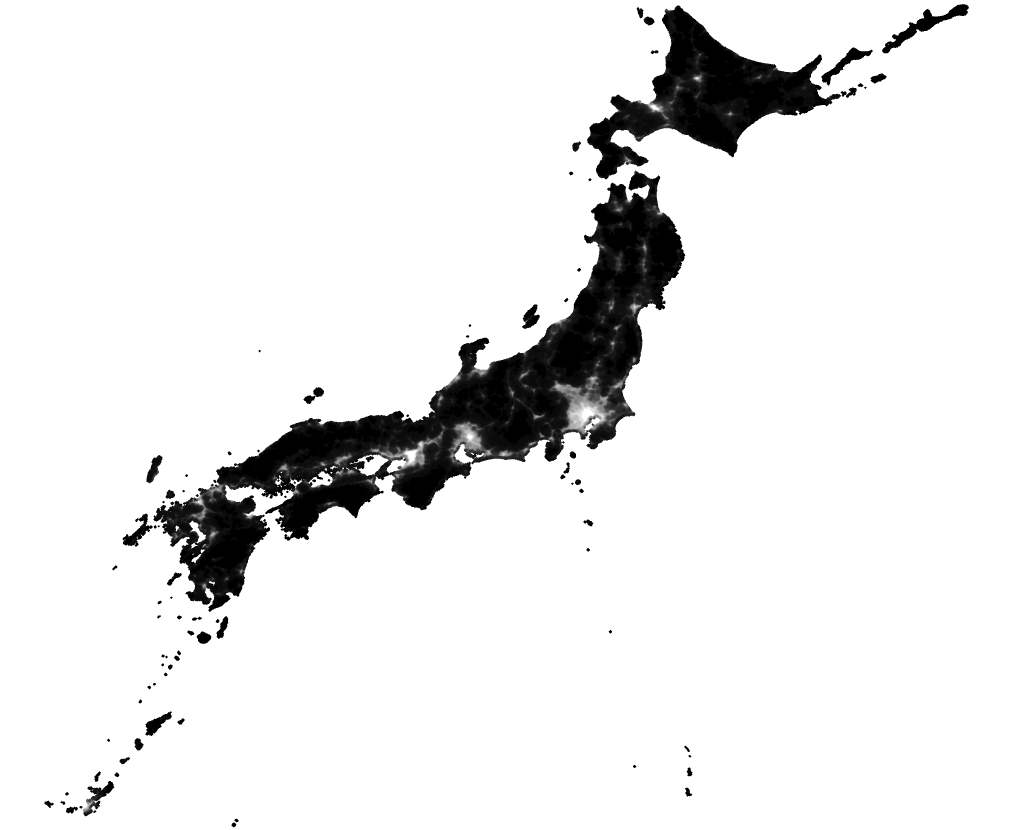}
\end{center}
\caption{Nighttime light distribution in Japan.}
\label{fig:jpnight}
\end{figure}

In Japan, "Regional Revitalization" has been prioritized by the Cabinet Secretariat to mitigate rural population decline and sustain regional economic vitality \cite{chihosousei}, placing domestic tourism at the core of regional growth strategies. However, evaluating tourism economies presents unique analytical challenges. Tourist demand is subject to complex temporal fluctuations influenced by seasonal weather patterns, regional institutional calendars (such as holiday periods), and origin-destination population flows.

Furthermore, as Kawamura (2008) \cite{kawamura} highlighted, tourism services represent non-storable commodities that cannot be produced in advance or held in inventory. Consequently, precise, high-frequency spatial tracking of tourist flows is far more critical for municipal planning and business operations than in traditional retail or manufacturing sectors. To address these societal needs, integrating satellite NTL monitoring with local statistics offers a promising methodology for capturing intra-municipal tourism dynamics and economic variations.

\subsection{Related Work}
The utility of Earth observation big data in economic research has been well documented. Donaldson et al. (2016) \cite{donaldson} underscored that modern petabyte-scale imagery access has bridged spatial data analytics with empirical economics and human geography. At the macroeconomic level, Henderson et al. (2012) \cite{henderson} proved that NTL emissions effectively capture sharp real-economy adjustments across international crises and localized economic events, bypassing the structural distortions and reporting delays inherent in official GDP metrics.

For urban and regional planning, Ichinose et al. (2002) \cite{ichinose} demonstrated that aggregating satellite data across administrative boundaries within Geographic Information Systems (GIS) provides a scalable substitute for missing or inconsistent municipal statistics. In developing contexts, Kurata (2017) \cite{kurata} verified that NTL data reflect diverse socio-economic conditions at district levels, including human capital and local living standards.

Regarding regional sustainability and economic spillovers, Chang et al. (2019) \cite{changli} showed that NTL metrics effectively evaluate regional Sustainable Development Indicators (SDIs) in China, filling statistical gaps and capturing regional spatial spillover effects. Focusing specifically on tourism dynamics, Chalkias et al. (2019) \cite{chalkias} analyzed European tourism hubs, demonstrating through linear and geographically weighted regression (GWR) that seasonal NTL variation strongly correlates with tourist arrivals and overnight stays. Their findings confirmed that nightlight observation provides a cost-effective, high-frequency tool for monitoring local seasonal economies—an approach directly relevant to municipal-level tourism analysis.

\section{Novelty and Utility}
\begin{figure}[h]
\begin{center}
\includegraphics[width=130mm]{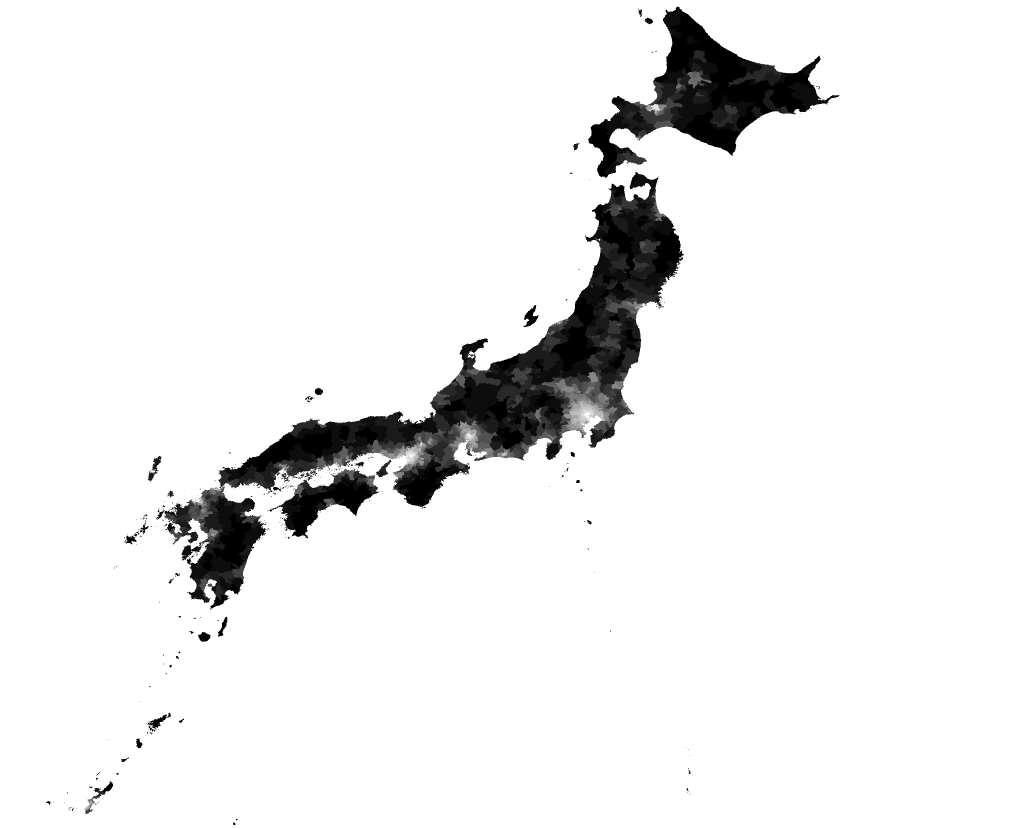}
\end{center}
\caption{Night light at Japan by city.}
\label{fig:jpnight_pref}
\end{figure}

This paper first describes the methodology of using nightlight data for remote sensing. Next, I examine the correlation between nightlight emissions and economic indicators within Japan, as well as the feasibility of utilizing nightlight data as a proxy variable. Furthermore, I discuss the spatial variations and characteristics of nightlight data. Finally, I demonstrate that a more detailed economic analysis can be achieved by analyzing the temporal transitions of nightlight data in conjunction with economic time-series data and existing macroeconomic data.

While numerous papers and books on the tourism industry have been published in Japan, no studies have yet reported on the relationship between domestic nightlight emissions and the tourism sector. Therefore, exploring the feasibility of using nightlight data as an economic indicator or proxy variable in the tourism industry constitutes a novel contribution of this study.

In particular, existing Japanese macroeconomic datasets are often aggregated at the municipality level as their minimum unit. Consequently, it is difficult to extract the economic vitality of finer sub-regions, such as town districts (cho-chome) or neighborhood boundaries (aza-kai), within the same municipality. Alternatively, while platforms like V-RESAS offer data with high temporal and spatial resolution, they only disclose data for specific areas, such as selected train stations, making them unsuitable for arbitrary or custom regional analyses.

For instance, in the case of Yuzawa Town, public data are aggregated and available for indicators such as population, GDP, the number of accommodation facilities, and tourist arrivals. However, from these existing macroeconomic indicators alone, it is difficult to discern that there are two major snow resorts within Yuzawa Town, let alone analyze each resort and its surrounding area.

To address this limitation, this study confirms that by supplementing existing macroeconomic indicators with nightlight data, I can identify the presence of two large snow resorts within the same municipality and capture the economic conditions of each respective area. In conclusion, the novelty and utility of this paper lie in demonstrating that the use of nightlight data enables a more detailed regional economic analysis in any arbitrary area than traditional macroeconomic indicators allow.

\section{Nightlight Data}
\subsection{Datasets Utilized}
Regarding the nightlight data, this study utilizes datasets acquired by the Defense Meteorological Satellite Program (DMSP), a United States Air Force meteorological satellite, consistent with the aforementioned prior research. The National Oceanic and Atmospheric Administration (NOAA) \cite{noaa1}, the publisher of the data, distributes several variations of the data commonly used as nightlight measurements, including radiance-calibrated data \cite{noaa2}, cloud-free compositions, and annual averages \cite{noaa3}.

From these available datasets, this study primarily employs the annual average nightlight data spanning from 1992 to 2013. The coverage area of the measurements ranges from -180 to 180 degrees longitude and -65 to 75 degrees latitude, with a spatial grid resolution of 30 arc-seconds. The period of the data used in this paper covers 1992 to 2013, during which the data were captured by DMSP satellite numbers 10 through 18 \cite{noaa3}.

\subsection{Raster Analysis}
Raster calculations and spatial analyses are performed on the acquired nightlight imagery using GIS software. For the vector layer, I utilize the "Administrative Boundary Data (National, 2020)" downloaded from the National Land Numerical Information download site of the Ministry of Land, Infrastructure, Transport and Tourism \cite{mlit}, or the "Statistics GIS (2015 Census, Small Area)" from the Portal Site of Official Statistics of Japan (e-Stat) \cite{estat, estatgis}, which integrates various socioeconomic statistical datasets. The acquired annual nightlight images are applied as the raster layer. By overlaying these layers, the nightlight intensity data are joined to each administrative boundary, making it possible to quantitatively compare them alongside various statistical data, as shown in \figref{fig:jpnight}\footnote{Cited from Otomo (2021) \cite{otomo}}. Furthermore, \figref{fig:jpnight_pref}\footnote{Cited from Otomo (2021) \cite{otomo}} visualizes the mean nightlight intensity calculated for each administrative district.

\section{Data-Driven Analysis of Yuzawa Town}
This chapter utilizes traditional macroeconomic data to examine the number of tourists and the socioeconomic time-series transitions of Yuzawa Town as a prominent snow resort.

\subsection{The Bubble Economy and the Resort Boom}

\begin{figure}[h]
\begin{center}
\includegraphics[width=130mm]{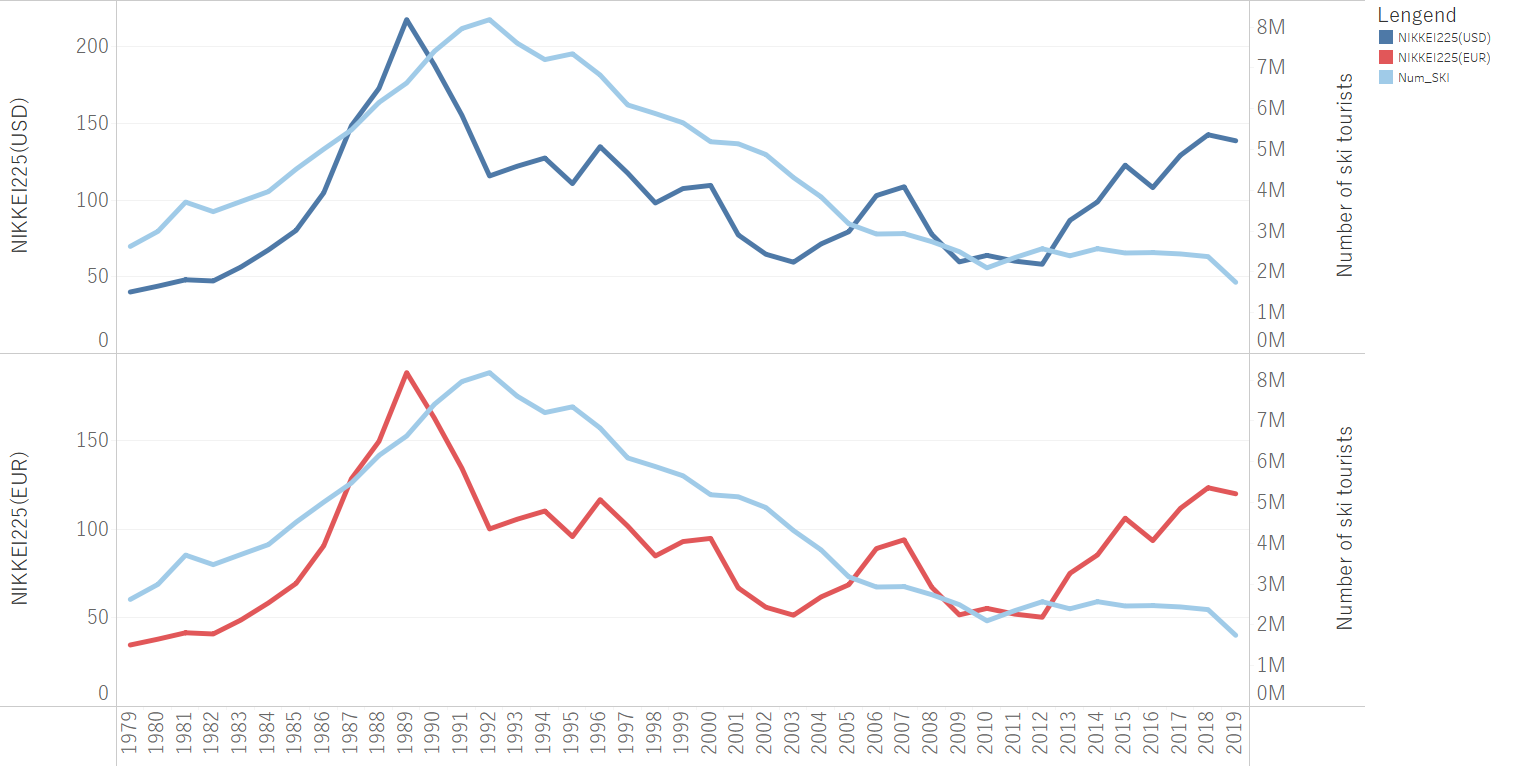}
\end{center}
\caption{Time-series graph of ski tourist arrivals and the Nikkei Stock Average}
\label{fig:ski_nikkei}
\end{figure}

Ski tourism in Japan experienced significant growth during the resort boom of the 1980s and early 1990s. Kureha (1995, 2002) \cite{kureha1995, kureha2002} drew substantial attention to this phenomenon, introducing the concept of a "ski tourism region" consisting of three core elements—the ski resort, the surrounding ski village, and ski tourists—and examined its developmental processes and regional conditions \cite{kankokenkyu}. According to tourism statistical data published by Yuzawa Town \cite{yuzawa_ski}, the number of ski tourists increased rapidly during the bubble economy period. As illustrated in \figref{fig:ski_nikkei}, the tourist numbers peaked and then turned into a sharp decline around 1992, slightly lagging behind 1989, the year when the annual average of the Nikkei Stock Average reached its historical highest value\footnotemark\footnotetext{The author calculated the annual averages based on daily data obtained from the reference source \cite{nikkei225}.}, or the collapse of the asset price bubble\footnotemark\footnotetext{In addition to stock prices, the economic downturn and public perception generally place the "bursting of the bubble" starting from 1991.}.

\subsection{The Collapse of the Bubble and Resort Condominiums}
According to Komenami (2000) \cite{komenami}, the resort condominium boom reached its peak during the expansionary phase of the bubble economy. Notably, Yuzawa Town accounted for more than one-third of the market, with 3,912 of the 11,564 resort condominium units sold nationwide in 1988 concentrated in this single municipality.

\begin{figure}
\begin{center}
\includegraphics[width=130mm]{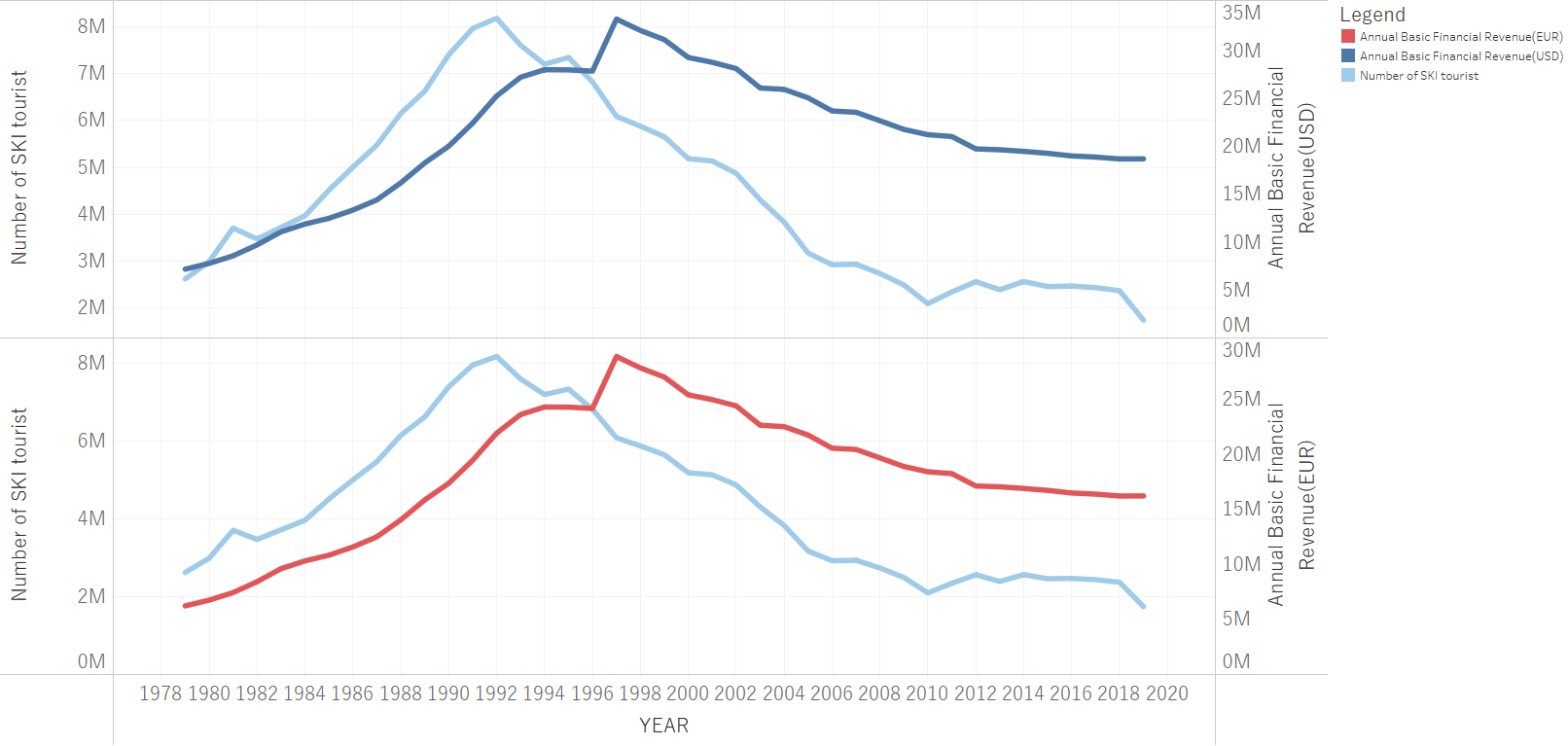}
\end{center}
\caption{Time-series graph of ski tourist arrivals and standard fiscal revenue}
\label{fig:ski_tax}
\end{figure}

Furthermore, an investigative reporting The Asahi Shimbun Reporting Team \cite{asahi_hudousan} noted that during the bubble era, owning a resort condominium was a symbol of wealth. However, following the collapse of the bubble, properties were dumped onto the market, causing prices to plummet. As owners stopped visiting, an increasing number of units failed to collect management fees, leading some properties to be forcibly liquidated through court-ordered auctions. The report also highlights that individuals who are delinquent on management fees are often equally delinquent on their fixed asset taxes. To examine this impact, I visualized the relationship between these factors and the number of ski tourists using local allocation tax datasets published by Yuzawa Town \cite{yuzawa_tax}. As illustrated in \figref{fig:ski_tax}, tax revenues began to decline with a slight time lag following the decrease in ski tourist arrivals.

\begin{figure}
\begin{center}
\includegraphics[width=130mm]{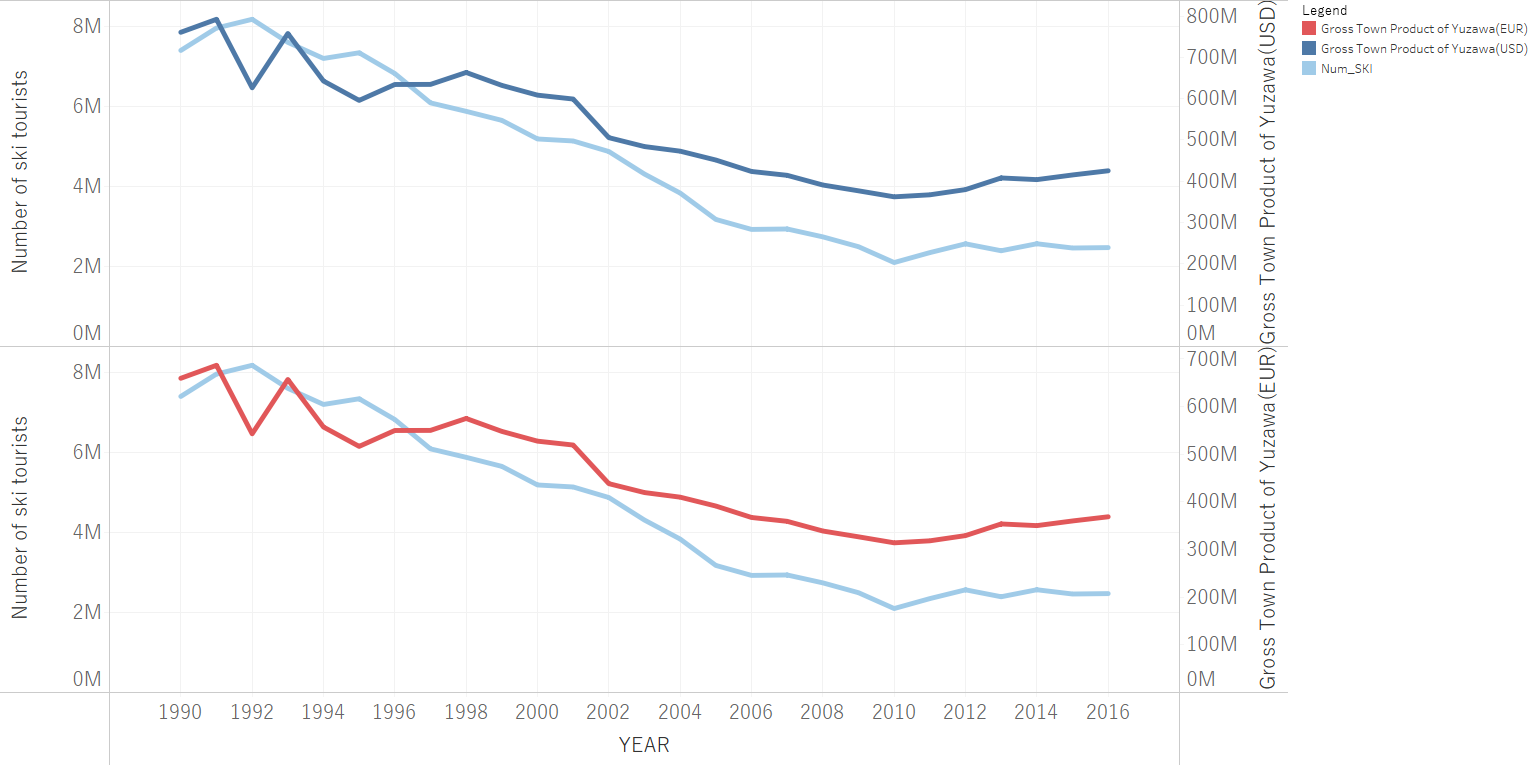}
\end{center}
\caption{Time-series graph of ski tourist arrivals and Yuzawa Town's Gross Municipal Product}
\label{fig:ski_sna}
\end{figure}

\subsection{Economic Impacts of Tourism}
According to Kawamura \cite{kawamura}, while production spillover effects lie at the core of economics, their realization process simultaneously generates income effects that increase value-added, employment effects that increase the number of employees, and tax revenue effects that boost public revenue.

In addition, a dialogue-format discussion in Development Strategies for Japanese-Style Resorts \cite{jpresort} describes the following dynamics: Yuzawa Town, which is renowned not only for skiing but also as a hot spring (onsen) resort, collected bath taxes (nyuto-zei) from both tourists and resort condominium owners, reaching an annual total of 120 million yen. The construction of these resort condominiums revitalized local supermarkets, retail stores, restaurants, and souvenir shops. In particular, because resort condominium owners seldom cooked for themselves, local restaurants frequently experienced long queues, especially on Sundays. These accounts vividly reflect the mechanism of local economic spillover effects.

\begin{figure}
\begin{center}
\includegraphics[width=130mm]{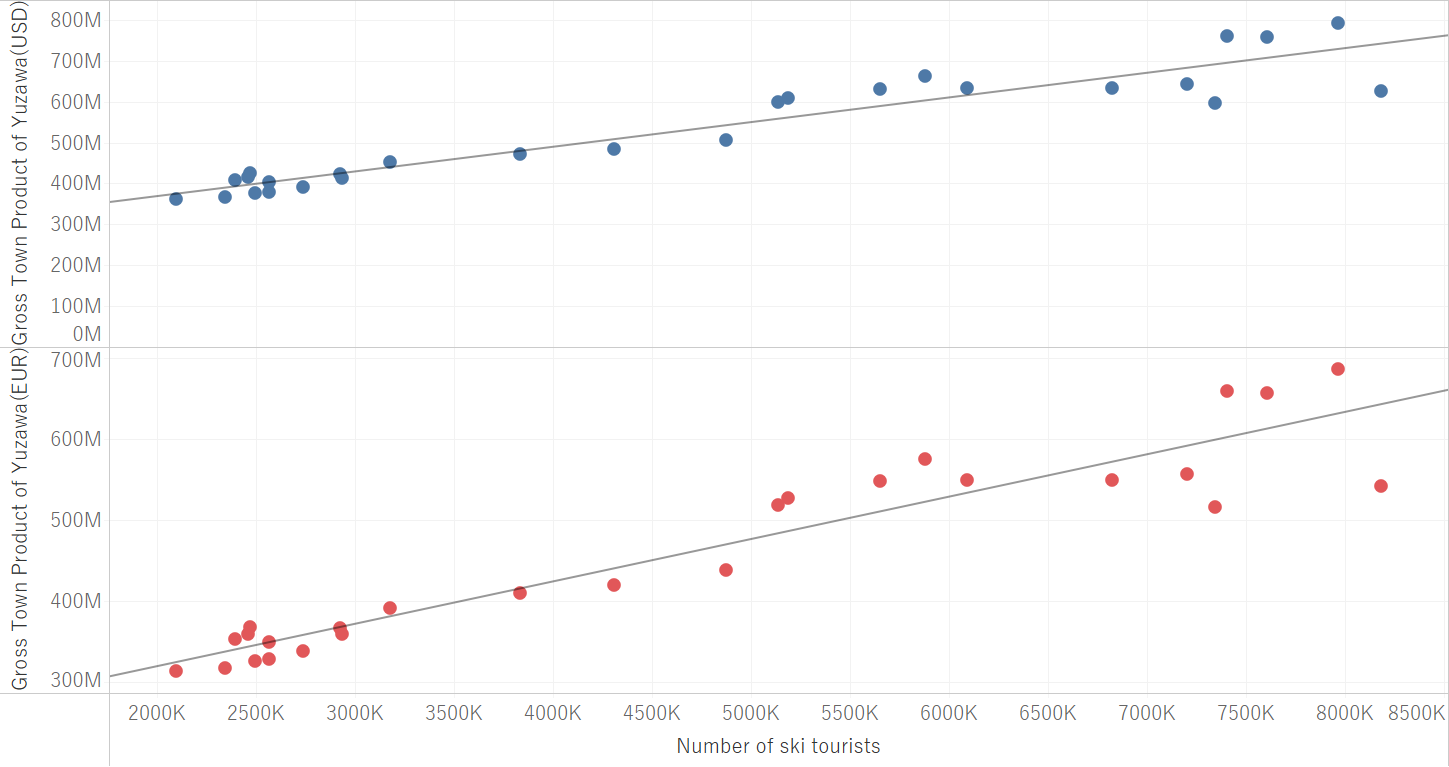}
\end{center}
\caption{Scatter plot illustrating the correlation between ski tourist arrivals and Yuzawa Town's Gross Municipal Product}
\label{fig:ski_gtp_scatter}
\end{figure}

\subsection{Yuzawa's Economy with Ski Tourist Arrivals as an Explanatory Variable}

Figures~\ref{fig:ski_sna} and~\ref{fig:ski_gtp_scatter} illustrate the relationship between ski tourist arrivals and the Gross Municipal Product (GMP) of Yuzawa Town, extracted from the Prefectural Accounts \cite{kenminkeizai} in the Statistical Yearbook of Niigata Prefecture \cite{toukeinenkan_nigata}.
Let $i$ ($1990 \le i \le 2016$, where $i$ is an integer) denote the year $G_i$ be the GMP of Yuzawa Town, and $S_i$ be the number of ski tourists.
All monetary values originally in JPY are converted to USD (and EUR) based on the Bank of Japan exchange rates as of August 3, 2026 ($1~\text{USD} = 156.50~\text{JPY}$ and $1~\text{EUR} = 180.69~\text{JPY}$) \cite{jpbank}.
The resulting relationship is expressed in thousand USD [or EUR] as follows:

\footnotetext{Although the original data are recorded in Japanese era calendar years, they were converted to Gregorian calendar years by the author. The same conversion applies to the data from \cite{yuzawa_ski, yuzawa_tax, toukeinenkan_nigata, kenminkeizai} used in this paper.}

\begin{equation}
G_i = 
\begin{cases}
603.5 S_i + 248,243 & \text{(Thousand USD)} \\
522.7 S_i + 214,954 & \text{(Thousand EUR)}
\end{cases}
\end{equation}

The above estimation yields $R^2 = 0.8916$ and $p < 0.0001$, indicating that ski tourist arrivals explain the Gross Municipal Product of Yuzawa Town with substantial explanatory power.

Next, spanning $1979 \le j \le 2019$ where $j$ is an integer, let $T_j$ be the standard fiscal revenue of Yuzawa Town and $S_j$ be the number of ski tourists. Classifying the period into pre-1990 ($j \le 1990$) and post-1990 ($j > 1990$) yields the following results in USD and EUR (in thousand USD and thousand EUR, respectively):

\begin{equation}
\text{USD: } 
  \left\{
    \begin{array}{ll}
      (j \le 1990), & T_j = 2.6767 S_j + 180.06 \\
      (j > 1990),   & T_j = 1.5157 S_j + 17,904.15
    \end{array}
  \right.
\end{equation}

\begin{equation}
\text{EUR: } 
  \left\{
    \begin{array}{ll}
      (j \le 1990), & T_j = 2.3183 S_j + 155.96 \\
      (j > 1990),   & T_j = 1.3128 S_j + 15,507.22
    \end{array}
  \right.
\end{equation}

The values of ($R^2$, $p$-value) before and after the collapse of the bubble economy are $(0.9715, 4.683 \times 10^{-9})$ and $(0.5113, 1.306 \times 10^{-5})$, respectively. Both equations demonstrate how representative ski tourist arrivals have been as a variable explaining the economic condition of Yuzawa Town. From these findings, although ski tourist arrivals do not directly correlate with nightlight emissions, it can be inferred that the tangible outcomes of their economic spillover effects—such as resort condominiums, hotels, and restaurants—collectively influence the nightlight intensity.

\section{Analysis and Discussion Using Nightlight Data}
This chapter first captures the characteristics of nightlight imagery. To facilitate a clearer understanding of these characteristics, I compare the nightlight patterns of a major metropolitan area with those of Yuzawa Town. Next, I verify the presence of two large snow resorts within Yuzawa Town that do not explicitly appear in traditional macroeconomic indicators. Finally, I discuss the relationship between Yuzawa Town's tourism data and nightlight emissions.
\subsection{Geospatial Comparison of Nightlight Data}
\begin{figure}[h]
\begin{center}
\includegraphics[width=130mm]{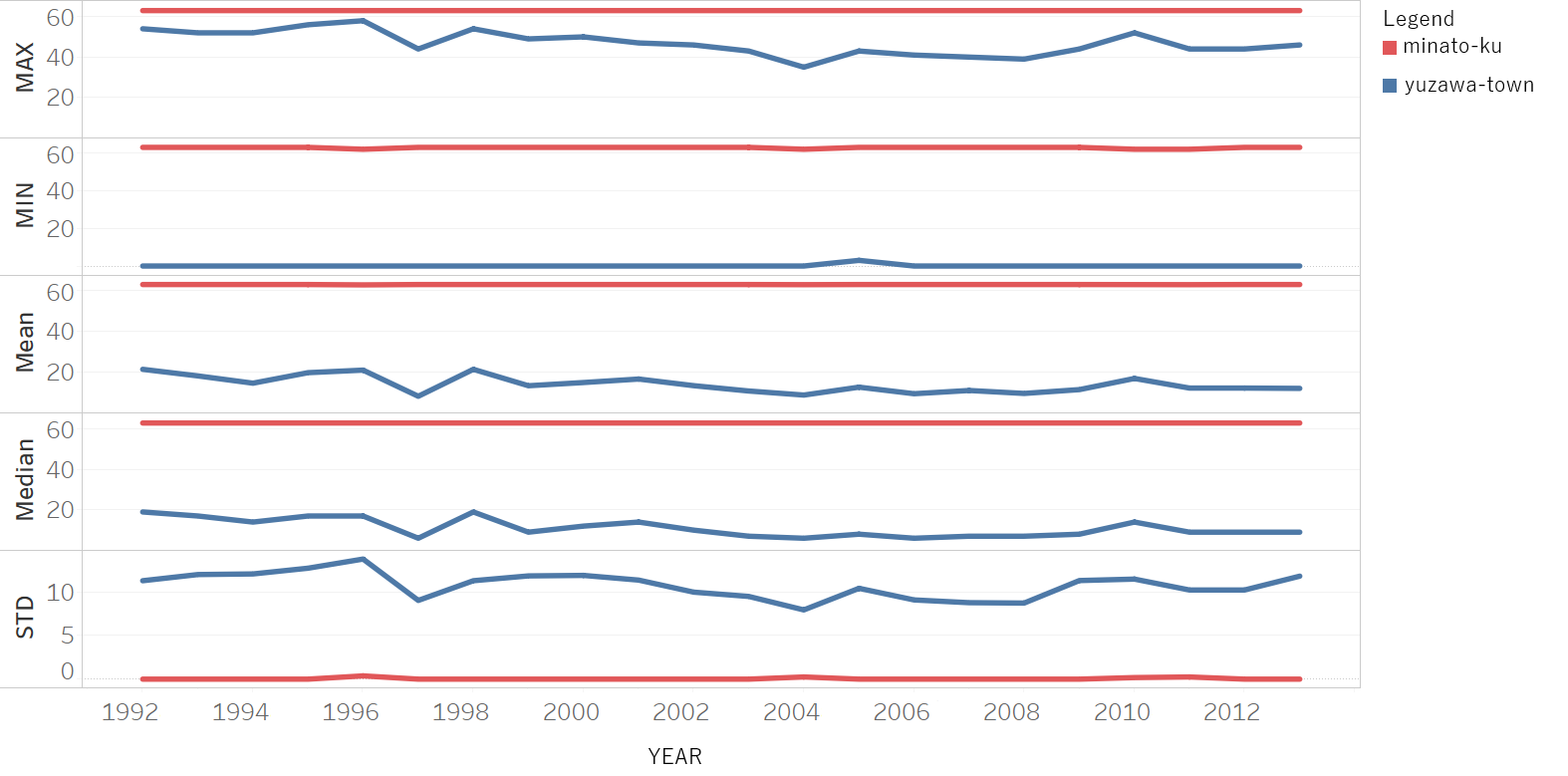}
\end{center}
\caption{Time-series graph of nightlight metrics for Minato Ward and Yuzawa Town}
\label{fig:nl_minato_yuzawa}
\end{figure}



\begin{figure}[htbp]
  \centering
  \begin{minipage}[t]{0.48\textwidth}
    \centering
    \includegraphics[width=\textwidth]{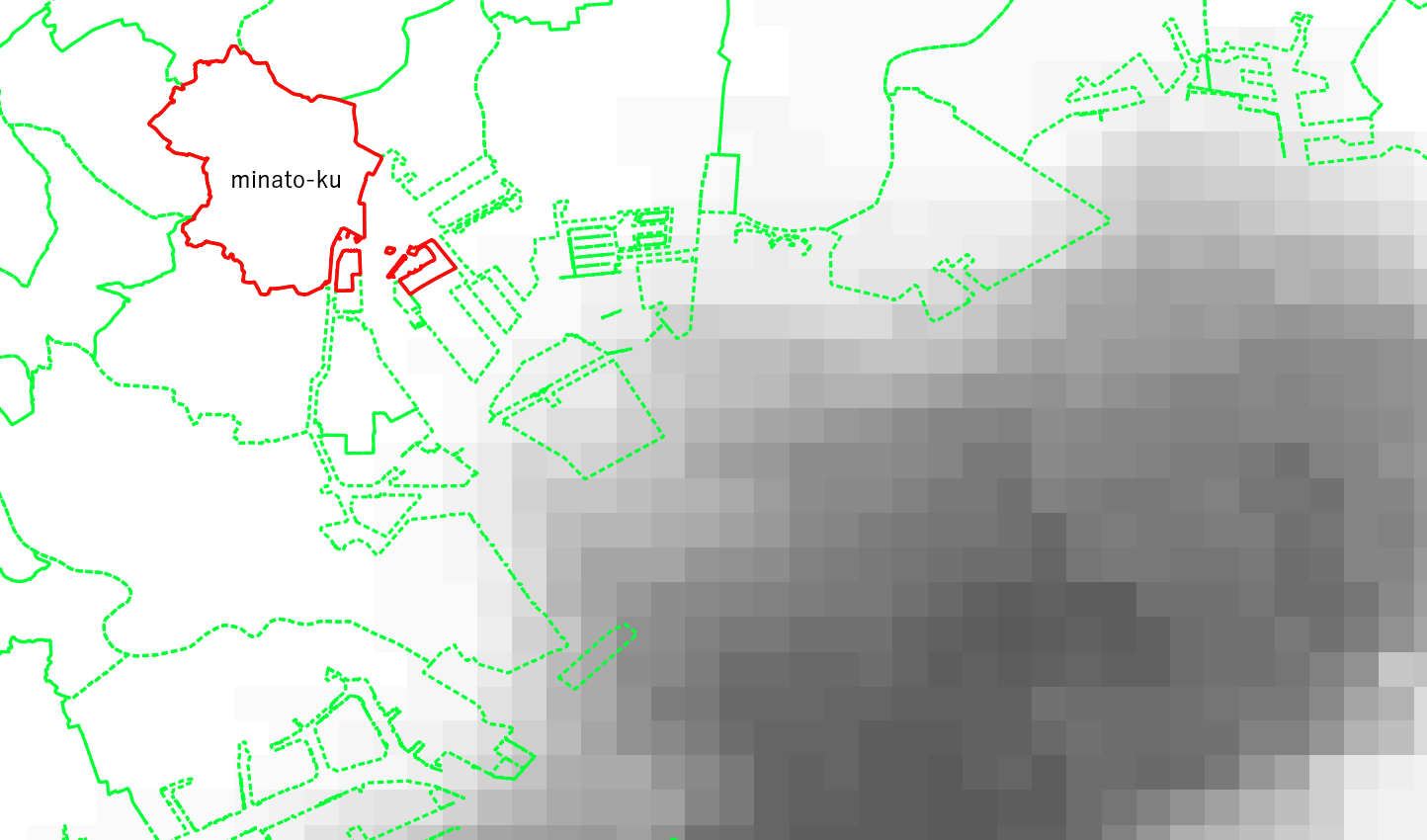}
    \caption{Nightlight emissions in Minato Ward (1992)}
    \label{fig:nl_minato_1992}
  \end{minipage}%
  \hfill
  \begin{minipage}[t]{0.48\textwidth}
    \centering
    \includegraphics[width=\textwidth]{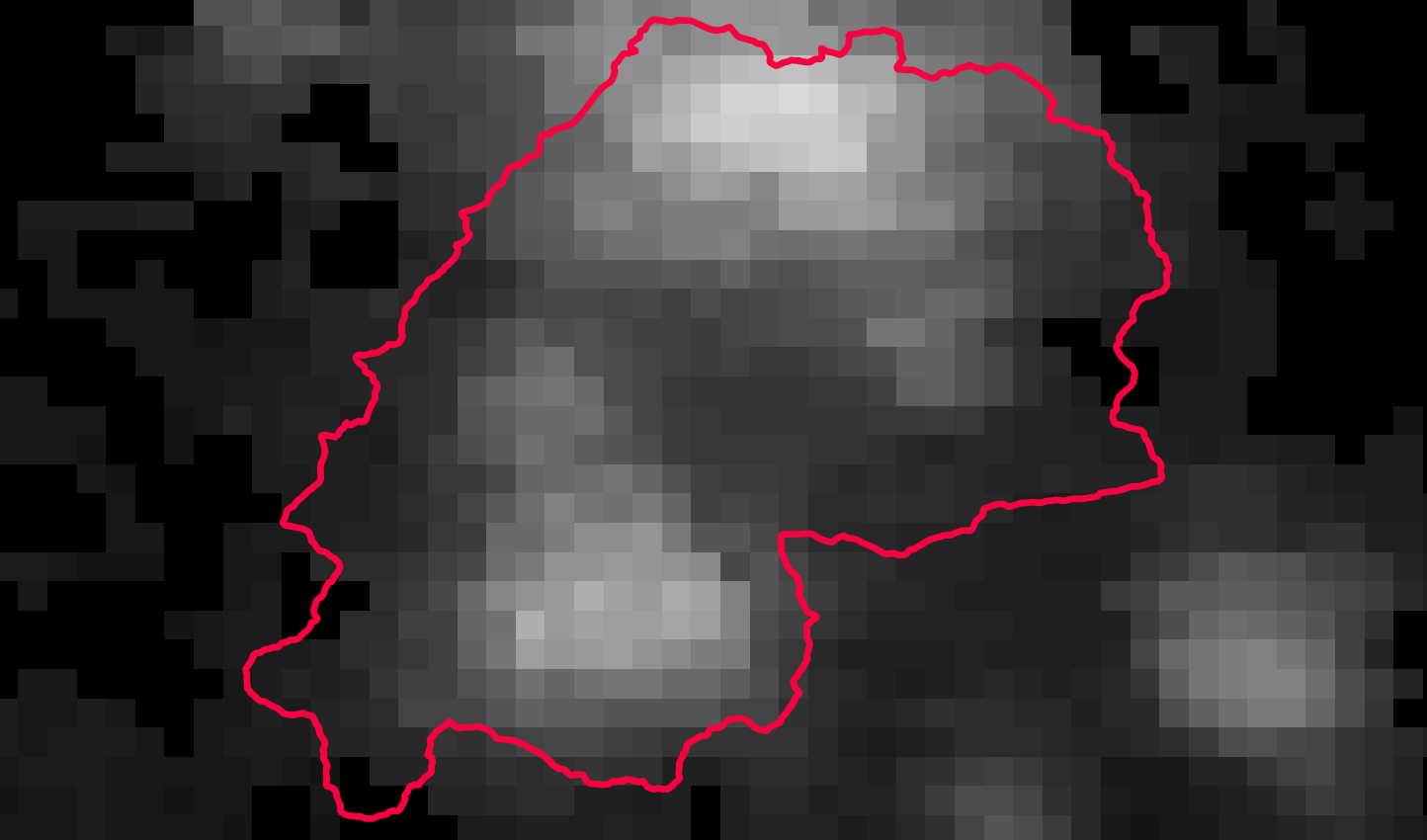}
    \caption{Nightlight emissions in Yuzawa Town (1992)}
    \label{fig:nl_yuzawa_1992}
  \end{minipage}
\end{figure}

To compare the nightlight data between Yuzawa Town and a so-called major metropolitan area, Minato Ward, Tokyo, is utilized. As shown in \figref{fig:nl_minato_yuzawa}, both the maximum (MAX) and minimum (MIN) nightlight values for Minato Ward from 1992 to 2013 consistently remain at nearly the maximum observable value
The same trend applies to the mean and median values. This indicates that in all sub-regions within the area, the captured nightlight values reached near the maximum observable threshold every year. Consequently, the standard deviation (STD) value remains nearly zero every year. Looking at \figref{fig:nl_minato_1992}, it can be observed that the entire area of Minato Ward—and the downtown core as a whole—appears mostly white, whereas the nightlight values over Tokyo Bay are discernibly darker.

In contrast, as illustrated in \figref{fig:nl_minato_yuzawa}, the maximum nightlight value for Yuzawa Town fluctuates depending on the year, while the minimum value remains nearly zero almost every year. Consequently, both the mean and median values exhibit annual variations, which are considered to reflect the economic vitality of Yuzawa Town in each respective year. Furthermore, the standard deviation values also vary annually, indicating a stark contrast between bright and dark areas within the region.



Examining \figref{fig:nl_yuzawa_1992} reveals that certain parts of the town are bright while others are dark, and that the illuminated areas within the same municipality of Yuzawa Town are largely divided into two distinct clusters.

\begin{figure}[h]
  \centering
  \begin{minipage}[t]{0.48\textwidth}
    \centering
    \includegraphics[width=\textwidth]{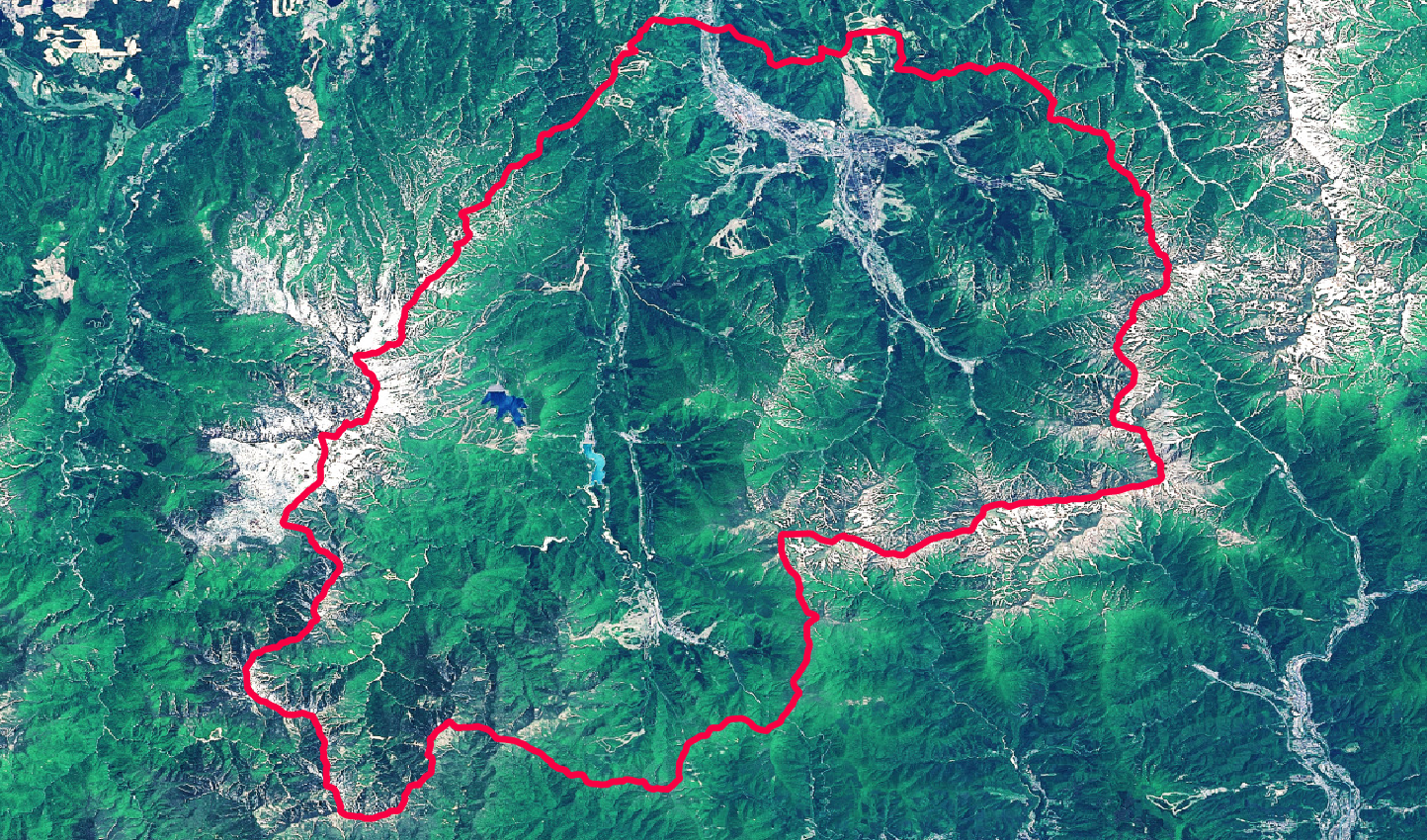}
    \caption{Photographic map of Yuzawa Town}
    \label{fig:sat_yuzawa}
  \end{minipage}%
  \hfill
  \begin{minipage}[t]{0.48\textwidth}
    \centering
    \includegraphics[width=\textwidth]{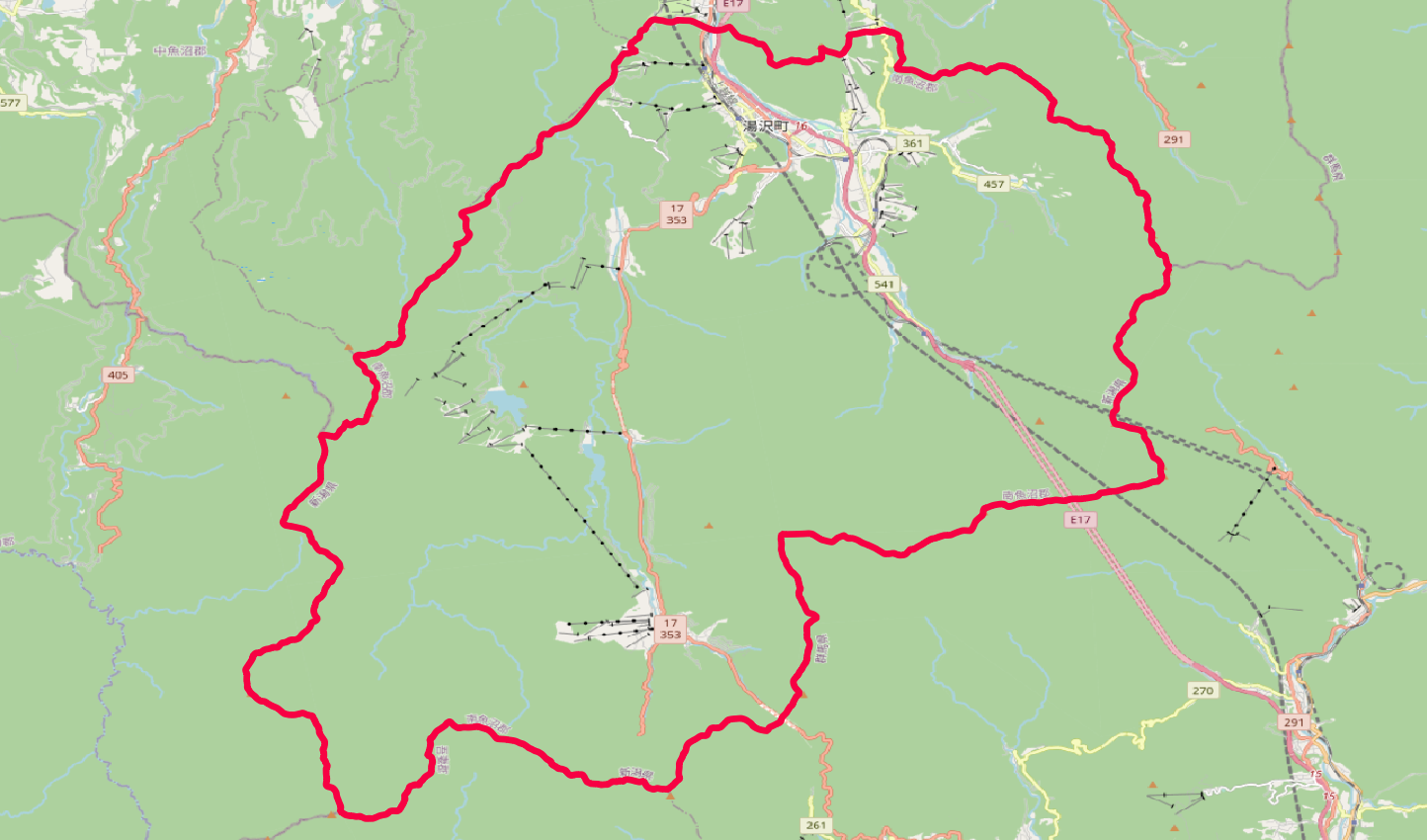}
    \caption{Yuzawa Town viewed via OpenStreetMap}
    \label{fig:osm_yuzawa}
  \end{minipage}
\end{figure}

\subsection{Regional Characteristics Viewed from Nightlight Data}
In nightlight imagery, white areas generally represent illuminated regions, and it can be inferred that brighter areas reflect more active human activities. Therefore, I examine the illuminated regions within Yuzawa Town.

When verifying the geography of Yuzawa Town using the Geospatial Information Authority of Japan (GSI) Maps \cite{chiriin}\figref{fig:sat_yuzawa}
, it can be observed that one of the bright regions expands around a large plain in the upper area, while the other is centered around a smaller plain in the lower area, and that the dark areas correspond to mountainous regions.

Next, when using OpenStreetMap \cite{osm} to analyze the geography of Yuzawa Town, \figref{fig:osm_yuzawa} reveals that the illuminated areas within the town correspond to two distinct clusters: first, the snow resorts adjacent to the area from GALA Yuzawa to the front of Echigo-Yuzawa Station where snow resorts are densely concentrated; and second, the area spanning from Naeba Ski Resort to Mitsumata/Kagura Snow Resort.

In other words, during the peak of ski tourist arrivals in 1992, each snow resort was bustling with numerous tourists visiting for skiing. Consequently, it can be confirmed that the illumination from tourism-related businesses, driven by economic spillover effects, was captured and reflected as nightlight emissions.

\subsection{Temporal Transitions of Yuzawa Town Viewed from Nightlight Data}

\begin{figure}
\begin{center}
\includegraphics[width=130mm]{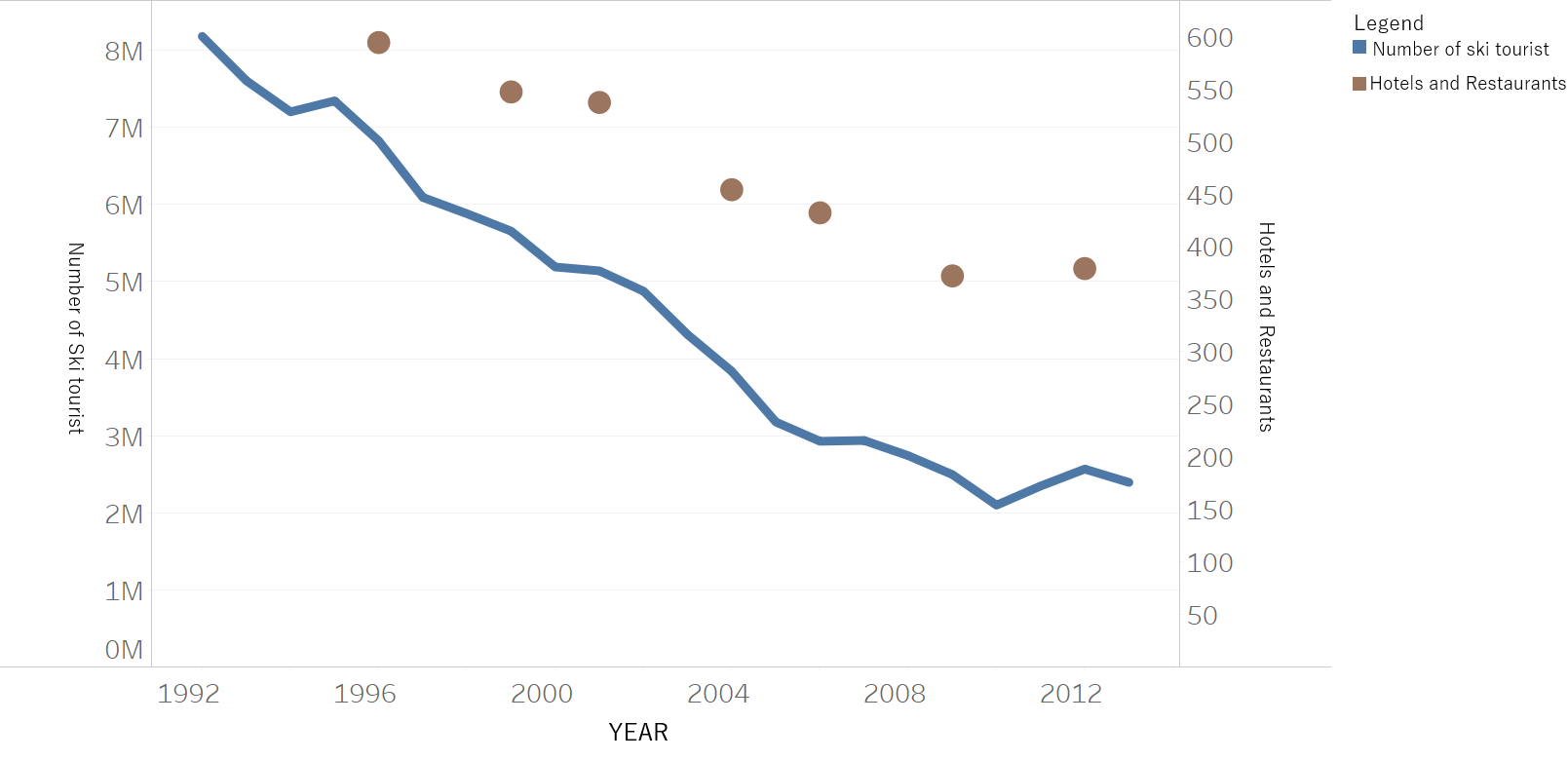}
\end{center}
\caption{Time-series of the number of ski tourists and the number of hotel and service providers.}
\label{fig:ski_hotels}
\end{figure}

The number of ski tourists in Yuzawa Town has been on a declining trend since its peak in 1992. In tandem with the decrease in ski tourist arrivals, the number of establishments in the hotel and tourism service industries has also declined. Utilizing data from e-Stat \cite{estat} and RESAS \cite{resas}, \figref{fig:ski_hotels}
illustrates the time-series graph constructed based on the number of establishments related to the hotel and tourism service industries.

\begin{figure}
\begin{center}
\includegraphics[width=130mm]{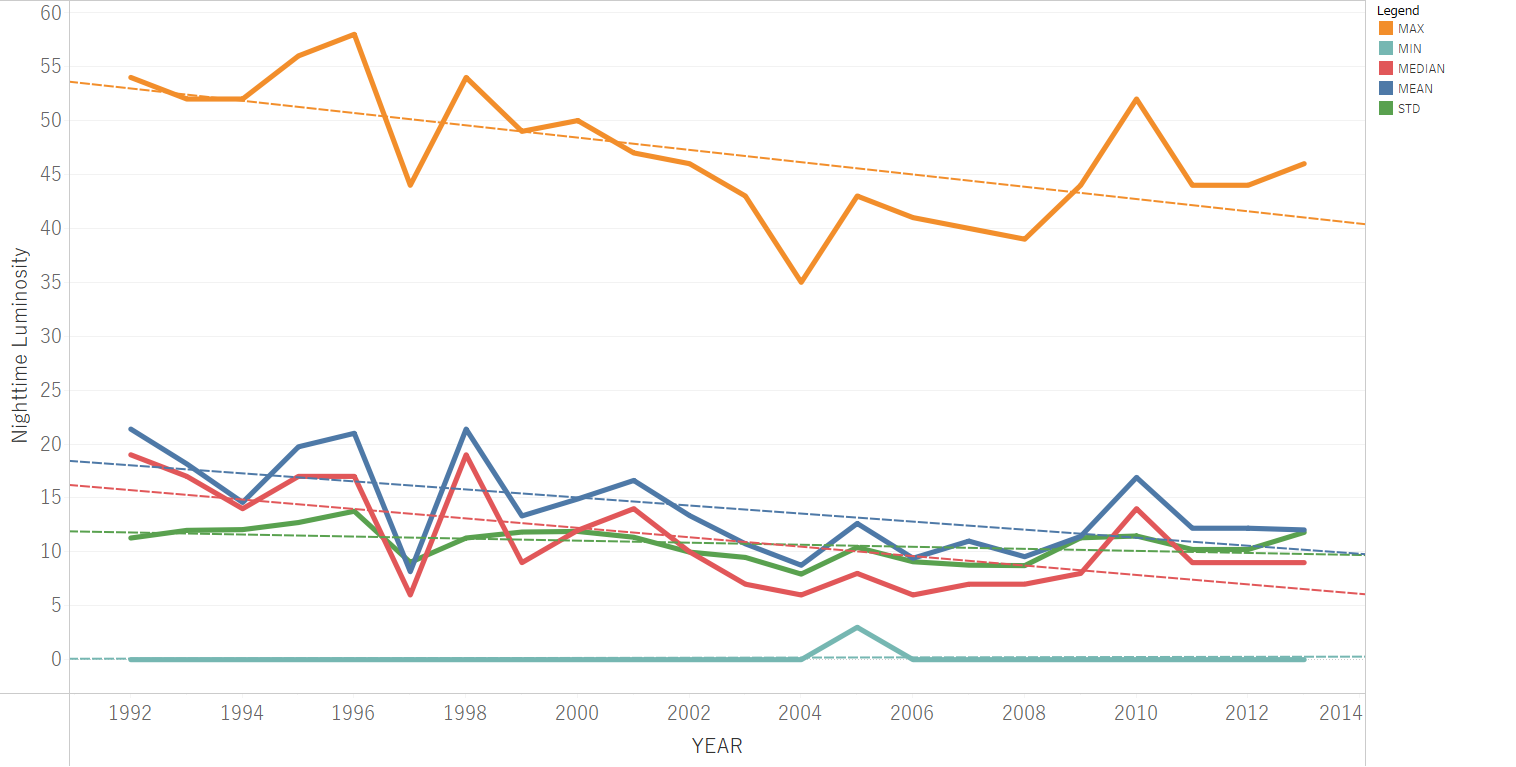}
\end{center}
\caption{Time-series of nighttime lights in Yuzawa Town}
\label{fig:ynl1992_2013ts}
\end{figure}

In association with this decrease in the number of establishments in the hotel and tourism service industries, the nightlight emissions diminish. Figure~\ref{fig:ynl1992_2013ts} illustrates this trend in a time-series graph. The maximum, mean, and median values of the nightlight emissions all exhibit a declining trend, and since the minimum value remains nearly zero almost every year, the standard deviation is also gradually decaying.

\subsection{Correlation Between Nightlight Data and Macroeconomic Indicators in Yuzawa Town}
\begin{figure}[h]
  \centering
  \includegraphics[width=\textwidth]{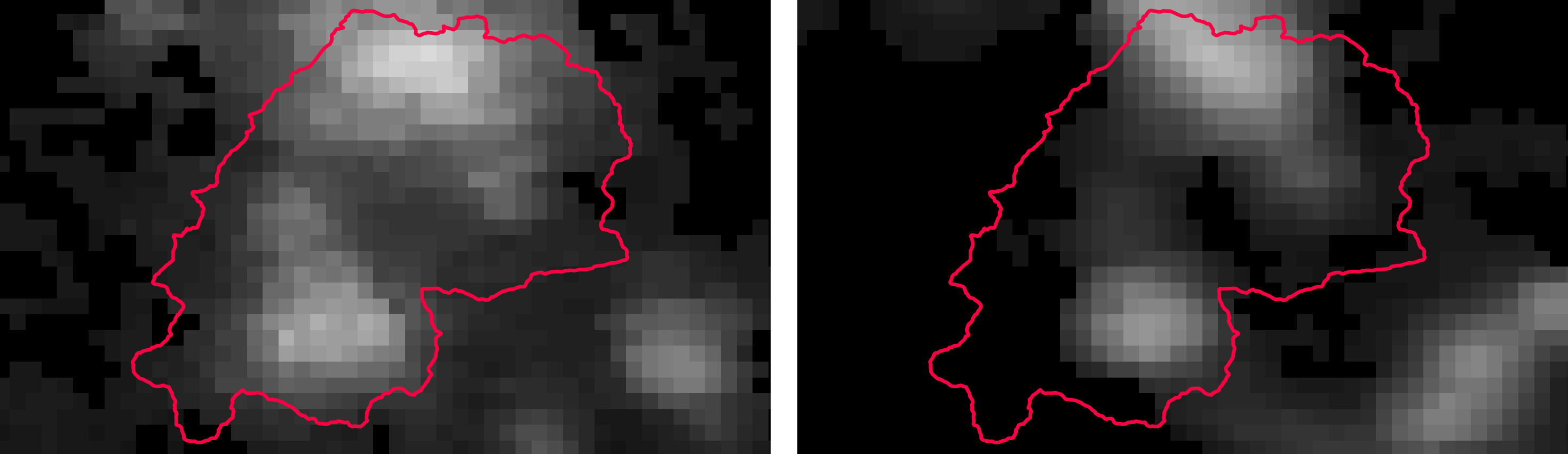}
  \caption{Comparison of nighttime lights in Yuzawa between 1992 and 2013}
  \label{fig:ynl1992_2013}
\end{figure}

\begin{figure}
\begin{center}
\includegraphics[width=130mm]{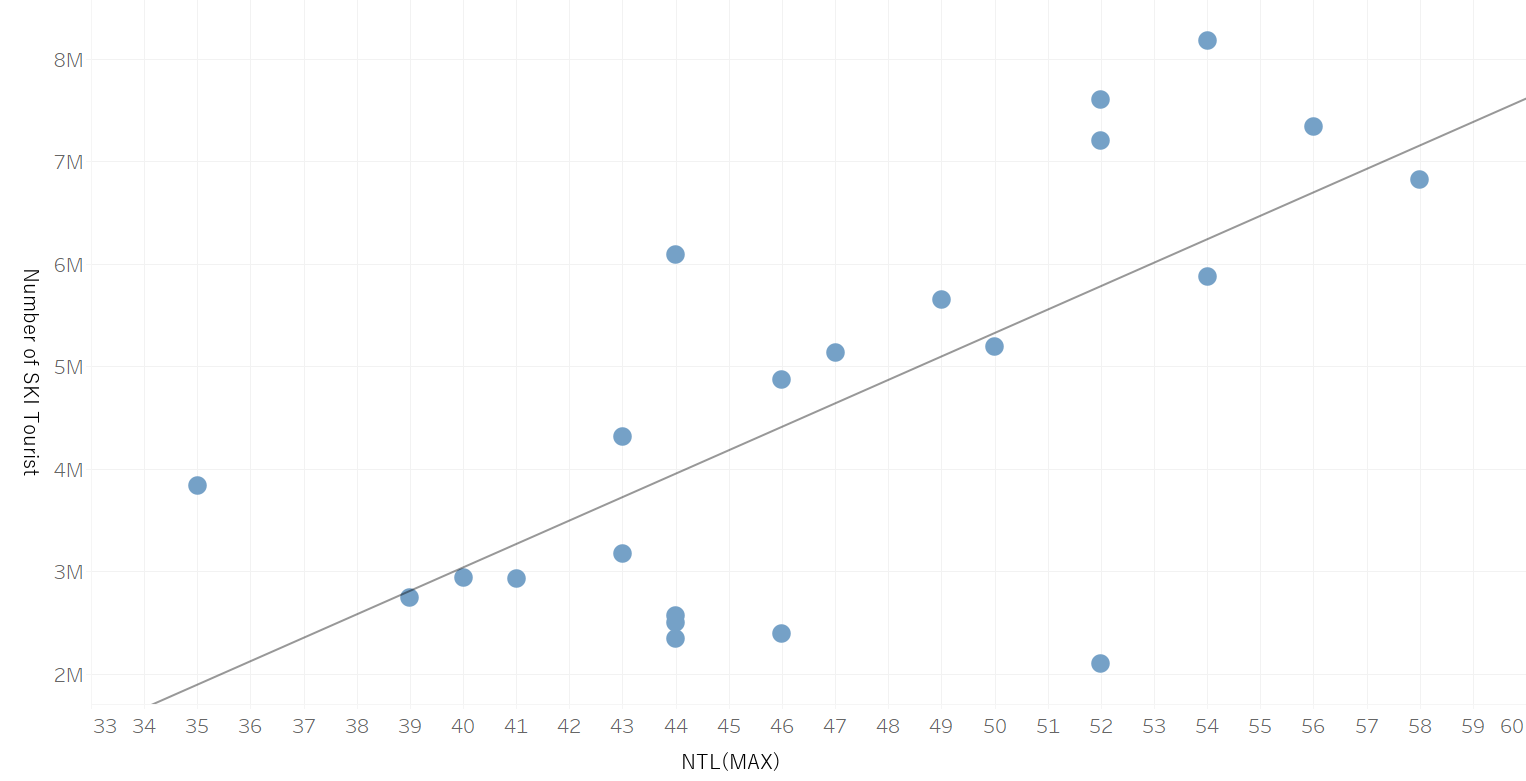}
\end{center}
\caption{Night Lights and Number of Ski Tourists}
\label{fig:nl_ski_scatter}
\end{figure}

\figref{fig:nl_ski_scatter} displays a scatter plot examining the extent to which nighttime light emissions decreased in conjunction with the decline in ski tourist arrivals.
Similarly, \figref{fig:ski_gtp_scatter} illustrates the relationship with Yuzawa Town's gross product.
Letting $i$ be an integer representing the years from 1992 to 2013 ($i = 1992, \dots, 2013$), $Y_i$ be the gross product of Yuzawa Town, and $N_i$ be the nighttime light emissions, the relationship is expressed by the following equation:

\begin{equation}
\label{eq:snow}
S_i \fallingdotseq 228791 \times N_i, \,R^2\fallingdotseq0.472, \, p\mathchar`-value < 0.001
\end{equation}

Consequently, within the scope of the analysis conducted using the acquired data, a certain degree of correlation was confirmed between the values obtained from the raster analysis of the nightlight imagery and the number of ski tourists in the case of Yuzawa Town, suggesting that quantitative evaluation is feasible.

\section{Conclusion}
While macro-level studies frequently highlight nighttime light (NTL) observation as a continuous proxy for cross-country GDP comparison (e.g., Henderson et al. \cite{henderson}; Kurata \cite{kurata}), this paper demonstrates a key counter-perspective: NTL imagery captures micro-level intra-municipal economic vitality that remains obscured within aggregated national statistics. As visually evident from the longitudinal analysis in Yuzawa Town (\figref{fig:ynl1992_2013}), NTL radiance directly reflects the temporal trajectory of local tourism, explicitly tracing the structural decline in ski tourist arrivals over past decades. Spatially reallocating commercial statistics (such as accommodation density) based on NTL intensity offers a fine-grained analytical approach superior to standard regional economic assessments.

Nevertheless, analytical limitations remain regarding dataset continuity. Free public archives from legacy NOAA sensors \cite{noaa2} do not cover post-2013 processing seamlessly without custom cross-sensor calibration. Although Yuzawa's NTL emissions display an overall downward trend, minor post-2008 fluctuations were observed. These short-term shifts may stem from inherent variance in annual compositing, or may reflect subtle micro-level responses to transient business expansions in 2009 and 2012. Aligning these trends with official establishment census data remains challenging due to mismatched reporting intervals. Resolving these ambiguities requires either long-term calibrated datasets or processing raw NASA Suomi-NPP VIIRS archives using consistent radiometric routines.

From a practical planning standpoint, integrating multi-modal Earth observation via platforms such as JAXA's G-Portal \cite{gportal} provides crucial analytical depth. For snow-resort economies like Yuzawa, sensor data from "SHIZUKU" (GCOM-W1) \cite{gcom_w}—which delivers high-frequency metrics on precipitation, soil moisture, and snow cover—can be coupled with NTL dynamics to build predictive models for weekly and seasonal tourist inflows. Additionally, vegetation metrics (NDVI/EVI) from "SHIKISAI" (GCOM-C) \cite{gcom_c} offer baseline environmental indicators for mid-to-long-term resort development and urban planning.

Future methodological refinements should incorporate time-series breakpoint analysis to detect structural economic shifts, alongside mesh-based spatial units to eliminate administrative boundary distortions. Finally, in light of severe disruptions caused by the COVID-19 pandemic on hospitality sectors, high-frequency satellite analytics offer essential policy value for rapid damage assessment and targeted administrative support (e.g., Sustainable Benefits disbursement) in tourism-dependent regions.

\paragraph{\texorpdfstring{Declaration of Generative AI in Scientific Writing}{Declaration of Generative AI in Scientific Writing}}
During the preparation of this work, the authors used ChatGPT-4 (OpenAI, accessed July 8, 2026) and Google Gemini (Google, accessed July 8, 2026) to enhance language clarity and readability. Specifically, these tools were used for \LaTeX{} formatting, including the embedding and placement of figures, table creation, and Bib\TeX{} reference adjustments.

\clearpage
\bibliographystyle{plainnat}
\bibliography{jssij2024_fixed_en}

\end{document}